\documentclass[12pt]{article}
\def\D{\Delta}
\def\d{\delta}
\def\L{\Lambda}
\def\l{\lambda}

\def\g{\gamma}
\def\e{\epsilon}
\def\s{\sigma}
\def\o{\omega}
\def\i{\iota}
\def\a{\alpha}
\def\b{\beta}
\def\cd{\cal D}
\def\cj{\cal J}
\def\cl{\cal L}
\def\m{\mu}
\def\th{\theta}
\def\f{\phi}

\def\rr{\bf R}
\def\ha{\frac12}
\def\dim{\textrm{dim}}
\def\det{\textrm{det}}
\newcommand{\be}{\begin{equation}}
\newcommand{\ee}{\end{equation}}
\newcommand{\bea}{\begin{eqnarray}}
\newcommand{\eea}{\end{eqnarray}}

\begin{document}
\begin{titlepage}

\bigskip
\bigskip
\begin{center}
\bf{\Large Tetrade Spin Foam Model}
\end{center}

\bigskip
\bigskip
\begin{center}
A. MIKOVI\'C\footnote{E-mail address: amikovic@ulusofona.pt}\\
Departamento de Matem\'atica,
Universidade Lus\'ofona de\\ Humanidades e Tecnologias,
Av. do Campo Grande, 376\\ 1749-024 Lisbon, Portugal
\end{center}

\bigskip
\bigskip
\centerline{{\bf Abstract}}

\bigskip
We propose a spin foam model of four-dimensional quantum gravity which is based on the integration of the tetrads in the path integral for the Palatini action of General Relativity. In the Euclidian gravity case we show that the model can be understood as a modification of the Barrett-Crane spin foam model. Fermionic matter can be coupled by using the path integral with sources for the tetrads and the spin connection, and the corresponding state sum is based on a spin foam where both the edges and the faces are colored independently with the irreducible representations of the spacetime rotations group. 

\end{titlepage}

\section{Introduction}

The approach of defining a quantum theory of gravity by using a path integral quantization has been revitalized by the appearance of the idea of spin foams \cite{b}. A spin foam model can be described as a lattice gauge theory for a BF theory, and although a BF theory is a topological theory, the Palatini action of General Relativity (GR) can be represented as a constrained BF theory, where the two-form $B$ is a wedge product of the spacetime tetrade one-forms. This then leads to the idea that the GR path integral could be defined as a modification of the path integral for a topological theory. This was the approach used for the construction of the Barret-Crane (BC) models \cite{bce,bc}, which culminated when a finite partition function for GR was constructed for any non-degenerate triangulation of the spacetime manifold \cite{cpr}. However, it was soon realized that one can obtain several finite BC models with different convergence properties \cite{baecr}. This ambiguity is a problem because it is still not clear which one of these has GR as the classical limit. The source of the ambiguity is the fact that the edge amplitudes of the dual two-complex cannot be fixed in the BC quantization procedure, which then leads to many possible models. 

Another problem with the BC type models is that it is difficult to couple matter, especially fermions, because matter fields couple to the tetrades and it is often impossible to rewrite the matter actions coupled to gravity as functionals of the $B$ and matter fields only. Several proposals for coupling matter have been made \cite{amm,cm}, but none of these proposals are based on manipulations of the matter plus gravity path integral, so that it is not clear what is their physical relevance.

These problems of the BC approach suggest that one should try to find a spin foam model which is based on the integration of the tetrade fields in the GR path integral. Such an approach should be feasible because the Palatini action is quadratic in the tetrads, so that the path integral over the tetrads is Gaussian. In this paper we show that the effective gauge theory obtained by the integration of the tetrads can be defined as a spin foam model. In section two we describe the construction of the path integral for pure gravity, while in section three we describe the construction of the path integral for gravity plus matter, including the cosmological constant term. In section four we present our conclusions. 

\section{Pure gravity path integral}

We consider first the case of pure gravity, and we start from the Palatini action 
\be S  =\int_M \e_{abcd}\, e^a \wedge e^b \wedge R^{cd} = \int_M \langle e^2  R \rangle \,d^4 x \quad,\ee
where $M$ is the spacetime manifold, $e^a$ are the tetrad one-forms, $R^{ab} = d\o^{ab} + \o^a_c \wedge\o^{cb}$ is the curvature two-form, $\o^{ab}$ is the spin connection one-form and $\e_{abcd}$ is the totally antisymmetric symbol ($\e_{0123}=1$). The corresponding path integral can be rewritten formally as
\be Z=\int {\cal D} \o \, {\cal D} e \,e^{i\int_M \langle e^2 R \rangle \,d^4 x} = \int {\cal D} \o \, (\det\, R )^{-1/2}\quad, \label{fpi}\ee 
where $(\det\, R)^{-1/2}$ denotes the result of the integration of the tetrads.

The formal expression (\ref{fpi}) suggests that one may try to define $Z$ on a triangulation of $M$ as
\be Z= \int \prod_l dA_l \prod_f (\det \, F_f )^{-1/2}= \int \prod_l dg_l \prod_f 
\D ( g_f)\quad,\ee
where $A_l =\int_l \o$, $g_l =e^{A_l}$, $\det \,F = (\e^{abcd}F_{ab}F_{cd})^2$,
\be g_f = e^{F_f}=\prod_{l\in\partial f}g_l \quad,\quad 
\D (g_f ) =(\det \, F_f )^{-1/2}\quad,\ee 
and the indices $l$ and $f$ stand for the edges and the faces of the dual two-complex of the triangulation. The group function $\D(g)$ should be gauge invariant, so that we take
\be \D (g) = \sum_\L \D(\L )\,\chi_\L (g) \quad,\ee
where $\chi_\L (g)$ is the character for an irreducible representation (irrep) $\L$, and the sum is over all irreps of a given category (finite-dimensional or unitary). This then implies that 
\be \D(\L)= \int_G dg \,\bar\chi_\L (g)\, \D(g)\quad.\label{pwi}\ee 

By using the formula
\be \int_G dg\, D^{(\L_1)\b_1}_{\a_1}(g)\cdots D^{(\L_4)\b_4}_{\a_4}(g) = \sum_\iota C^{\L_1\cdots\L_4 (\iota)}_{\a_1 \cdots \a_4}\left(C^{\L_1\cdots\L_4 (\iota)}_{\b_1 \cdots \b_4}\right)^* \quad, \label{fi}\ee
where $C^{\L_1\cdots\L_4 (\iota)}_{\a_1 \cdots \a_4}$ are the components of the intertwiners $\iota$ for the tensor product of four irreps and $D^{(\L)}(g)$ are the corresponding representation matrices, we will obtain a state sum of the form
\be Z= \sum_{\L_f,\iota_l} \prod_f \D(\L_f) \prod_v A_v
(\L_f,\iota_l) \quad,\label{grss}\ee
where the vertex amplitude $A_v$ is given by the evaluation of the pentagon spin network, which in the $SU(2)$ case is known as the $15j$ symbol. The state sum (\ref{grss}) is of the same form as in the case of the topological theory given by the BF action; however, the weights we put on the faces are not $\dim\,\L_f$ but the functions $\D(\L_f)$.

These new weights are given by the integrals which are generically divergent, due to 
$\det\, F_f  = 0$ configurations, so that some kind of regularization must be used. In order to do this, let us write the Lie algebra element $F_f$ as
\be F_f = \vec E_f \cdot \vec K + \vec B_f \cdot \vec J \quad,\ee
where $\vec K$ are the boost generators, while $\vec J$ are the spatial rotations generators. The $so(4)$ Lie algebra is a direct sum of two $so(3)$ algebras, and a basis of this decomposition is given by 
\be \vec {\cj}_\pm = \frac12 ( \vec J  \pm \xi\vec K ) \quad,\label{lrd}\ee
where $\xi =1$ in the Euclidian case and $\xi = i$ in the Minkowski case.

From (\ref{lrd}) it follows that 
\be \det\, F = (\vec E \cdot \vec B )^2 = \frac{\xi^2}{16}((\vec E_+ )^2 - 
(\vec E_- )^2 ) \quad,\ee
where $\vec E_\pm = \vec B \pm \frac{1}{\xi}\vec E$. Note that $\vec E_\pm$ are real in the Euclidian case, while in the Minkowski case are complex conjugates ($\bar E_+ = E_-$). The group function $\D(g_f)$ is then given by the expression
\be \D(g)= \frac{4}{\xi}((\vec E_+ )^2 - (\vec E_- )^2 )^{-1} \quad,\quad g=e^{\vec E_+ \cdot \vec{\cj}_+ + \vec E_- \cdot \vec{\cj}_-} =g_+ \,g_-\quad.\label{gfe}\ee

Let $E_\pm = |\vec E_\pm |$, then the weights $\D(\L_f)$ can be calculated from (\ref{pwi}) and (\ref{gfe}). Since $D^{(\L)}(g)$ is a direct product of two $SU(2)$ representations $D^{(j_+)}(g_+)$ and $D^{(j_-)}(g_-)$, we obtain
\bea \D(\L) &=& \int_{{\rr}^3}{d^3 \vec E_+ \over 8\pi^2}{\sin^2(E_+ /2)\over E_+^2} \int_{{\rr}^3}{d^3 \vec E_- \over 8\pi^2}{\sin^2(E_- /2)\over E_-^2}\nonumber\\&\,& {\sin(j_+ +\ha)E_+\over \sin(E_+/2)}\, {\sin(j_- +\ha)E_-\over \sin(E_-/2)}
\,\frac{4}{\xi(E_+^2 - E_-^2)} \quad.\label{eint}\eea
We have used the $SU(2)$ group integration measure
\be \int_{SU(2)}dg\,f(g) = \int_{{\rr}^3}{d^3 \vec E \over 8\pi^2}\,{\sin^2(E /2)\over E^2}\,f(E) \quad,\ee 
where $g=e^{\vec E \cdot \vec \s}$ is a spin $\ha$ representation matrix \cite{fk}, and the trace formula
\be \chi^{(j)}(g) = {\sin(j+\ha)E\over\sin(E/2)}\quad.\ee

Since $d^3 \vec E = E^2 \, \sin\th\, dE\, d\th\, d\f$, and there is no angular dependence in the integral (\ref{eint}), we can put $d^3 \vec E = 4\pi E^2 dE$, so that
\be \xi\pi^2 \D(\L ) = \int_0^{\infty} dE_+ \m_+ (E_+) \int_0^{\infty} {dE_- \m_- (E_-)\over  E_+^2 - E_-^2 } \quad,\ee
where $\m_\pm (E) = \sin (E/2) \sin (j_\pm + \ha )E$. The integral
\be I=\int_0^{+\infty} {dE_- \m_- (E_-)\over  E_+^2 - E_-^2 } \ee
is divergent, and has to be regularized. By using
\be \sin (E/2) \sin \left(j + \ha \right)E = \cos (jE) - \cos (j+1)E \quad,\ee
we obtain $I= I_j - I_{j+1}$, where
\be I_k = \int_0^{\infty} {dE \cos (kE)\over  E_+^2 - E^2 }\quad. \ee
Since
\be I_k = \ha \,Re\,\int_{-\infty}^{\infty} {dE \,e^{ikE}\over  E_+^2 - E^2 }\quad, \ee
it has a form of a one-dimensional quantum field theory (QFT) propagator. One can then regularize $I$ by the QFT methods, i.e. by replacing $E_+^2$ by $E_+^2 + i\e$ ($\e>0$) and by performing a contour integral along the real axis and a semi-circle in the upper half-plane \cite{ram}. After taking the $\e\to 0$ limit, one obtains
\be I_k = -\pi {\sin (k E_+ )\over 2E_+} \quad,\ee
so that
\be -\xi\pi \D(j,l) = \int_0^{\infty}{dE\over 2E}[\cos (j E) - \cos (j + 1)E ][\sin l E - \sin (l + 1)E] \quad.\label{a}\ee

In order to evaluate (\ref{a}) we will study the integral
\be I(z) = \int_\e^\infty {dE\over E}\, e^{-zE} \,,\quad z\in {\bf C}\,,\quad \e>0 \,.\ee
This integral can be rewritten as
\bea I(z) &=& \int_\e^\infty {dE\over E}\, e^{-zE}\\
             &=& \int_1^\infty {dx\over x}\,e^{-z\e x}= E_1 (z\e) \quad.\eea
Since
\be E_1 (z) = -\g -\log z - \sum_{n=1}^\infty {(-1)^n z^n \over n n!}\quad,\quad |arg\, z| <\pi \quad,\ee
where $\g$ is the Euler-Mascheroni constant \cite{as}, then
\be I(z) = -\ln\e -\g - \ln |z| - i\, arg\,z + O(\e) \quad,\quad |arg\, z| <\pi \quad.\ee
Let 
\be I(j,l)=\int_0^\infty {dE\over E}\cos (jE) \sin (l E) \quad,\ee
then
\bea I(j,l) &=& \lim_{\e\to 0}\int_\e^\infty {dE\over 2E} [\sin (j+l)E - \sin (j-l)E ]\\
&=& \ha \lim_{\e\to 0}\,Im\,[E_1 (-i(j+l)\e) - E_1 (i(l-j)\e)]\\ &=& \frac14 \pi + \ha arg\,(i(l-j)) = \frac{\pi}{2} \theta (l-j) \quad.\eea 
Therefore
\be \D(j,l) = -{1\over 2\xi} [2\th (l-j ) - \th (l -j +1)  -\th(l-j-1 )] \quad.\label{pgd}\ee

This implies that the non-zero coeficients are the ones with $l-j=0$ or $l-j=\pm 1$, whose values are given by
\be \D(j,j-1) = {\th_0\over 2\xi} \quad,\quad \D(j,j) = {1-2\th_0 \over 2\xi} \quad,\quad \D(j,j+1) = {\th_0 -1\over 2\xi}\quad,\ee
where $\th_0 = \th (0)$. The natural values for $\th (0)$ are $0$ and $1$, although $\th (0) =\ha$ gives a more symmetrical
$\D (j,l)$. In any case, one obtains a weight which is concentrated around the simple irreps $(j,j)$, so that the model can be considered as a generalization of the Barrett-Crane model.

\section{Path integral with sources}

Including matter and the cosmological constant term requires the evaluation of the path integral
\be Z=\int {\cd}e {\cd}\o {\cd}\psi \exp\left( i \int_M \left(\langle e^2 R \rangle +\l\langle e^4 \rangle\right)d^4 x + iS_m [\psi ,e,\o ] \right) \quad,\label{polpi}\ee
where $\l$ is the cosmological constant, $S_m = \int_M d^4 x \,{\cl}_m$ and ${\cl}_m$ is a function of the tetrads, spin connection, matter fields $\psi$ and their derivatives. Note that in the case of spin-half fermions ${\cl}_m$ is  a polynomial in $e$ and $\o$ given by
\be S_m = \int_M \e_{abcd}\,e^a \wedge e^b \wedge e^c \wedge \bar\psi \left(\g^d \left( d + \ha\o_{rs}\g^r\g^s \right)+ me^d \right)\psi \quad,\ee
where $\g^a$ are the Dirac gamma matrices and $m$ is the fermion mass. Hence the path integral (\ref{polpi}) can be evaluated at least perturbetively. 

In the case of scalar and vector bosons ${\cl}_m$ is not a polynomial function of the tetrads. However, in the vector boson case, when the Yang-Mills action is put on the simplicial complex, it is possible to obtain a polynomial approximation suitable for the spin foam evaluation \cite{amym}. 
 
When ${\cl}_m$ is a polynomial of the fields and their derivatives, the path integral (\ref{polpi}) can be evaluated perturbatively via
\be Z= \lim_{j,J,{\chi}\to 0} e^{i\l\int_M \langle (\d/\d j)^4 \rangle \,d^4 x + iS_m [-i\d/\d {\chi} ,-i\d/\d j ,-i\d/\d J ]} Z_0 [j,J,\chi ] \quad,\ee 
where
\be  Z_0 [J,j,{\chi}]=\int {\cal D}e \, {\cal D}\o \,{\cal D}\psi \, e^{ i\int_M  \left(\langle e^2 R \rangle+ J_{ab}\o^{ab} +j_a e^a + {\chi}^\a \psi_\a  \right)\,d^4 x} \label{gfm}\ee
is the generating functional. $Z_0$ is essentially the gravitational path integral with the sources since the matter integration in (\ref{gfm}) gives a delta function $\d(\chi)$. The tetrade path integral is Gaussian, so that we need to define
\be Z_0 [J,j] = \int {\cal D} \o \, e^{i\int_M d^4 x \,J_{ab}\o^{ab}} (\det R )^{-1/2} e^{-i\int_M d^4 x\int_M  d^4 y 
\langle j(x) R^{-1}(x,y)j(y) \rangle /4}\quad. \label{pifs}\ee

This expression can be defined on a triangulation of $M$ along the lines of the $J=j=0$ case. However, when the sources are present a more intricate state sum will appear. Guided by the expression (\ref{pifs}) we will define $Z_0$ as
\be Z_0 (J,j)= \int \prod_l dg_l \,\m(g_l ,J_l )\prod_f \D(g_f ,j_\e ) \quad,\label{gis}\ee
where
\be \m(g_l ,J_l ) = e^{iTr(\o_l J_l )} \quad,\quad \D(g_f ,j_\e )= \D(g_f) e^{-i \langle j_\e F_f^{-1} j_{\tilde\e}\rangle/4 } \quad,\ee
and the subscripts $\e$ and $\tilde\e$ denote two edges of the triangle dual to the face $f$. 

The function $\m$ is not a gauge invariant, and therefore it will be expanded as
\be\m (g_l,J_l) = \sum_{\L_l} \m (\L_l ,J_l) D^{(\L_l)}(g_l) \quad.\ee
Hence the coefficients $\m(\L,J)$ are matrices given by 
\be \m(\L,J)=\int_G dg\, \bar{D}^{\L}(g)\,\m (g,J)\quad. \label{propi}\ee
The function $\D$ will be expanded as a gauge invariant function, so that 
\be \D = \sum_{\L_f} \D (\L_f ,j_\e) \chi_{\L_f }(g_f)\quad.\ee
Hence
\be \D(\L,j)= \int_G dg \,\bar\chi_\L (g)\, \D(g,j)\quad.\label{dsi}\ee 

The group integrations in (\ref{gis}) can be performed by using the analog of the formula (\ref{fi}) for the tensor product of five irreps. One then obtains a state sum
\be Z_0(J,j)=\sum_{\L_f , \L_l  ,\i_l} \prod_f \D (\L_f , j_\e)  
\prod_l \m (\L_l ,J_l) \prod_v A_v (\L_f , \L_l, \i_l ) \quad.\ee
This is a novel spin foam state sum, because it involves a dual 2-complex whose edges and faces are independently colored with the irreps of the group. $Z_0(J,j)$ can be understood as an amplitude for a Faynman diagram given by a five-valent graph whose edges carry the irreps $\L_l$ and the loops carry the irreps $\L_f$. The edges have propagators $\m (\L_l ,J_l)$ and each loop carries a weight $\D (\L_f, j_\e)$, while the vertex amplitudes $A_v$ are given by the evaluation of the pentagon spin network with five external edges, where the internal edges carry $\L_f$ irreps, while the external edges carry $\L_l$ irreps.

The propagators $\m(\L,J)$ will be given by the integrals (\ref{propi}), 
where $\m(g,J)=e^{i(\vec\o_+ \cdot \vec J_+ + \vec\o_- \cdot\vec J_- )}$, $g=e^{\vec\o_+ \cdot\vec{\cj}_+ +\vec\o_- \cdot\vec{\cj}_-}$ and $\L=(k,l)$. In order to see what kind of integrals are these and what kind of regularization should be used, we will evaluate their traces, which are given by
\be Tr\,\m(\L,J) = \int_G dg \,Tr\,\m (g,J)\, \bar\chi_{\L}(g)\quad.\ee
We then obtain
\bea Tr\,\m(k,l,J_\pm) &=& \int_0^\infty {d\o_+ \over 4\pi}\m_+ (\o_+)\int d\th_+ \sin\th_+ e^{i\o_+ J_+ \cos\th_+}\nonumber\\
&\,&\int_0^\infty {d\o_- \over 4\pi}\m_- (\o_-)\int_0^\pi d\th_- \sin\th_- e^{i\o_- J_- \cos\th_-}\quad.\label{egw}\eea
This integral can be easily evaluated by using the formulas from the previous section, so that we obtain
\be Tr\,\m (k,l, J_\pm ) = {\th(J_+ - k)-\th(J_+ - k -1)\over J_+}\,{\th(J_l - l)-\th(J_- - l -1)\over J_-}\quad.\label{naw}\ee

Note that this function is non-zero only for the spins belonging to the intervals $[J_\pm -1 , J_\pm ]$, so that the sum over the $\L_l$ irreps is a finite sum. However, 
the weights (\ref{naw}) are not analytic functions in the vicinity of $J=0$, so that in the limit $J\to 0$ the partition function will diverge. In order to regularize this, we will calculate (\ref{egw}) perturbatively in $J$ by expanding
$e^{i\o J \cos\th}$ in the Taylor series around $J=0$. The $\m$ will be then a power series in $J$, with the coefficients given by the integrals
\be c_n (k)= {i^n \over n!}\int_0^\infty d\o \sin (\o/2) \sin (k+\ha)\o \,(\o J)^n \int_0^\pi d\th \sin\th \cos^n \th \quad.\ee
After performing the $\th$ integration, we obtain
\be c_n (k) = {i^n \over (n+\ha)n!}\int_0^\infty d\o [\cos (k\o) - \cos (k+1)\o ]\o^n \quad,\ee
where $n$ is an even number. This integral can be regularized by performing an analytic continuation of
\be \int_0^\infty e^{-\a\o}\,\o^n d\o = n!\, \a^{-n-1}\quad,\quad Re\,\a >0 \quad,\ee
to $Re\,\a =0$. This gives
\be c_{2n}(k) = {i\over n+\ha} [(k+1)^{-2n-1}-k^{-2n-1}] \quad,\ee
so that
\be Tr\,\m (k,l,J_\pm )= \sum_{m,n \ge 0} c_{2n}(k)\, c_{2m}(l) J_+^{2m}J_-^{2n} \quad.\label{muex}\ee 

The weights $\D(\L,j)$ are the character expansion coefficients of the group function 
\be \D (g,j) = (\det\,F )^{-\ha}\, \exp \left(-i {j_a  j_b^{\prime} \e^{abcd}F_{cd}\over 2\sqrt{\det\,F}} \right)\quad,\ee
where we have used $F^{-1}=-\ha F^* (\det\,F)^{-1/2}$.
By using the formulas from the previous section, we obtain
\be \D (g,j)= (E_+^2 - E_-^2 )^{-1} \exp\left(i\,{\vec Q_+ \cdot\vec E_+ + \vec Q_- \cdot\vec E_- \over E_+^2 - E_-^2 }\right) \quad,\ee
where $Q_\pm$ are quadratic functions of $j$ and $j^\prime$. The weights $\D$ will be then given by the integral
\bea \D &=& \int_0^\infty {dE_+ \over 4\pi}\m_+ (E_+)\int d\th_+ \sin\th_+ \nonumber\\
&\,&\int_0^\infty {dE_- \over 4\pi}\m_- (E_-)\int_0^\pi d\th_- \sin\th_- 
{\exp\left( i{Q_+ E_+ \cos\th_+ + Q_- E_- \cos\th_- \over E_+^2 - E_-^2}\right)\over E_+^2 - E_-^2 }\,.\eea

This integral can be evaluated perturbatively in $Q_\pm$, which is sufficient for our purposes, since at the end of calculation of the partition function the sources $j$ have to be set to zero. One can use the Taylor series
\bea &\,&\exp\left( i {Q_+ E_+ \cos\th_+ + Q_- E_- \cos\th_- \over E_+^2 - E_-^2}\right)\nonumber\\
 &=& \sum_{n=0}^\infty \frac{i^n}{n!}\left( {Q_+ E_+ \cos\th_+ + Q_- E_- \cos\th_- \over E_+^2 - E_-^2}\right)^n \quad,\eea
and because of the angular integrations only the even powers of $n$ contribute. The corresponding integrals can be regularized by the same procedure as in the previous section, i.e. by replacing $(E_+^2 -E_-^2 )^{-1}$ with $(E_+^2 -E_-^2 + i\e )^{-1}$, $\e >0$, integrating along the contour in the upper half-plane and then taking the $\e\to 0$ limit.

We then obtain
\bea \D (\L ,j) &=& \D (\L ) + \a_+(\L)Q_+^2 + \a_-(\L) Q_-^2 \nonumber\\ 
&+& \b_+(\L) Q_+^4 + \b_-(\L)Q_-^4 + \g(\L) Q_+^2 Q_-^2 + O(Q^6) \quad,\label{dex}\eea
where $\D(\L)$ is given by the equation (\ref{pgd}). For example
\bea \a_\pm (\L) &=& -\ha\int_0^\infty {dE_+ \over 4\pi}\m_+ (E_+)\int d\th_+ \sin\th_+ \nonumber\\
&\,&\int_0^\infty {dE_- \over 4\pi}\m_- (E_-)\int_0^\pi d\th_- \sin\th_- 
{ E_\pm^2 \cos^2\th_\pm \over (E_+^2 - E_-^2 +i\e )^3}\quad.\eea
It is easy to see that $\a_+ (k,l) = -\a_-(l,k)$ and one can show that $\a_- (k,l)$ is given by a convergent integral
\be \a_- (k,l) = - {2\over 3\cdot 16^2 \pi}\int_0^\infty dE \sin(E/2)\sin((k+1/2)E)[f_l(E) -f_{l+1}(E)],\ee
where
\be f_l (E) = -{l\over E^2} \cos(lE) + {l^2 E + 1\over E^3}\sin(lE) \quad.\ee 

Note that in the expansion (\ref{dex}), as well as in the expansion (\ref{muex}), only the even powers of the $\pm$ components appear, which means that the $SO(4)$ invariance is preserved. More explicitely
\be J_\pm^2 = \ha J^{ab} J_{ab} \pm \e^{abcd}J_{ab}J_{cd} \quad,\quad Q_\pm^2 = \ha f^{ab} f_{ab} \pm \e^{abcd}f_{ab}f_{cd}\quad,\ee
where $f_{ab} = \ha (j_{a}j^\prime_{b}-j_{b}j^\prime_{a})$. Since $f_{ab}$ is associated to a triangle, there will be three source vectors $j_1$, $j_2$ and $j_3$ for each edge of the triangle, and therefore we will take a symmetric expression
\be f_{ab} = \frac13 \left( f_{ab}^{(12)}+f_{ab}^{(23)}+f_{ab}^{(31)} \right) \quad.\ee

\section{Conclusions}

We have formulated an approach to obtain a spin foam model of quantum gravity based on the integration of the tetrade fields in the path integral for the Palatini action for General Relativity. We have determined the weights for the spin foam state sum in the Euclidian gravity case, and one can see that the tetrade model is a generalization of the BC model since all the irreps of $SO(4)$ are included in the state sum but the weights are non-zero only for the irreps close to the simple irreps. Beside allowing for the coupling of the fermionic matter, the tetrade model is an improvement over the BC model since all the simplex weights can be determined (in the BC case the edge amplitudes are arbitrary).  

Matter and cosmological constant can be coupled by introducing the sources for the tetrades and the spin connection, which results in a new type of the spin foam state sum. It is a sum of the amplitudes for the spacetime two-complex where the edges and the faces carry independent labels. If one thinks about a spin foam as a world-sheet of a spin network, then including sources corresponds to putting additional spins on the verticies of the spin network. This is close to the picture where matter corresponds to the external edges of an open spin network, an approach which was used in \cite{amm} to couple matter. However, the formalism of \cite{amm} requires open spin foams, while for the tetrade model only closed spin foams are used. 

In order to have a physical model, the simplex weights should be defined for the Minkowski case. One way to do this would be to perform an analytic continuation of the Euclidian weights (\ref{pgd}) such that $\xi \to i$. In this approach one would work with the same category of representations as in the Euclidian case, i.e. finite-dimensional $SU(2,{\bf C})\times SU(2,{\bf C})$ representations, so that the Minkowski weights will be the Euclidian weights times the appropriate factors of $i$. An alternative approach would be to use the category of unitary $SL(2,{\bf C})$ representations. In this case the irreps are infinite-dimensional and can be labeled as $(j,\rho)$ where $2j\in\bf {\bf Z}_+$ and $\rho\in{\bf R}_+$. The corresponding weights will be determined by the integrals (\ref{pwi}), (\ref{propi}) and (\ref{dsi}). 

An important next step is to study the convergence of the state sum in the Euclidian and the Minkowski case. Note that if the Euclidian state sum turns out to be divergent, it can be regularized by passing to the quantum group at a root of unity, which is usually done in the case of topological spin foam models. However, since our model is non-topological, using a quantum group regularization is not necessary, so that one can use alternative regularizations, for example a gauge fixing procedure for spin foams \cite{fr}.

As far as the semiclassical limit is concerned, this is still an unsolved problem for quantum gravity spin foam models. The difficulty is that in the case of non-topological models the partition function state sum is triangulation dependent. This means that it is not a diffeomorphism invariant, as it should be. Formally one can sum over all the triangulations in order to obtain a triangulation independent expression; however, it is still not known how to do this. Perhaps adapting the techniques of the dynamical triangulations approach \cite{ajl} could help with this problem. 

An alternative approach to the problem of finding the semiclassical limit would be to find a smooth manifold limit. This would require studying triangulations with very large number of simplexes, and hopefully one could extract an effective diffeomorphism invariant action. However, a technique must be developed in order to do this, perhaps something analogous to the methods used for the study of the two-dimensional Ising model near the critical point.  

The simplest thing one can do in order to better understand the model is to analyze the state sum for the simplest triangulation of the four-sphere given by the hexagon graph\footnote{The four-sphere can be triangulated by six four-simplices, so that the dual one-complex is the five-valent hexagon graph consisting of six verticies and 15 edges.}. Clearly, more work is necessary in order to resolve these issues, but the model is promising because it was obtained from the discretized path integral for the Palatini action.

\bigskip
\noindent{\bf ACKNOWLEDGEMENTS}

This work was partially supported by the FCT grant POCTI/MAT/45306 /2002.

\end{document}